\documentclass[aps,prl,twocolumn,floatfix,preprintnumbers,altaffilletter,superscriptaddress, nofootinbib]{revtex4-1}
\usepackage{graphicx,url,amssymb,amsmath,rotating,color,units,wasysym,epsfig,multirow,epstopdf,subcaption, comment}
\usepackage[colorlinks,urlcolor=blue,citecolor=blue,linkcolor=blue]{hyperref}
\usepackage{soul,xcolor}

\begin{document}
\title{Measuring the primordial gravitational-wave background\\
in the presence of astrophysical foregrounds}

\author{Sylvia Biscoveanu}
\email{sbisco@mit.edu}
\affiliation{LIGO, Massachusetts Institute of Technology, Cambridge, Massachusetts 02139, USA}
\affiliation{School of Physics and Astronomy, Monash University, VIC 3800, Australia}
\affiliation{OzGrav: The ARC Centre of Excellence for Gravitational-Wave Discovery, Clayton, VIC 3800, Australia}

\author{Colm Talbot}
\affiliation{LIGO, California Institute of Technology, Pasadena, CA 91125, USA}
\affiliation{School of Physics and Astronomy, Monash University, VIC 3800, Australia}
\affiliation{OzGrav: The ARC Centre of Excellence for Gravitational-Wave Discovery, Clayton, VIC 3800, Australia}

\author{Eric Thrane}
\affiliation{School of Physics and Astronomy, Monash University, VIC 3800, Australia}
\affiliation{OzGrav: The ARC Centre of Excellence for Gravitational-Wave Discovery, Clayton, VIC 3800, Australia}

\author{Rory Smith}
\affiliation{School of Physics and Astronomy, Monash University, VIC 3800, Australia}
\affiliation{OzGrav: The ARC Centre of Excellence for Gravitational-Wave Discovery, Clayton, VIC 3800, Australia}

\setstcolor{blue}
\begin{abstract}
Primordial gravitational waves are expected to create a stochastic background encoding information about the early Universe that may not be accessible by other means.
However, the primordial background is obscured by an astrophysical foreground consisting of gravitational waves from compact binaries.
We demonstrate a Bayesian method for estimating the primordial background in the presence of an astrophysical foreground. 
Since the background and foreground signal parameters are estimated simultaneously, there is no subtraction step, and therefore we avoid astrophysical contamination of the primordial measurement, sometimes referred to as ``residuals.''
Additionally, since we include the non-Gaussianity of the astrophysical foreground in our model, this method represents the statistically optimal approach to the simultaneous detection of a multi-component stochastic background.
\end{abstract}

\maketitle

{\em Introduction.}---Detection of a cosmological gravitational-wave background from the early Universe is one of the most ambitious goals of gravitational-wave astronomy.
There are several scenarios which may give rise to primordial backgrounds, including inflationary scenarios and phase-transition scenarios~\cite{maggiore}.
Inflationary models in general produce a gravitational-wave background through the amplification of vacuum fluctuations~\cite{inflation, Turner:1996ck, Easther:2006vd, Guzzetti:2016mkm}.
In the slow-roll inflation model, the dimensionless energy density of the background,
\begin{align}
\Omega_\text{gw}(f) \equiv \frac{1}{\rho_c}
\frac{d \rho_\text{gw}}{d \ln f} ,
\end{align}
is expected to be $\Omega_\text{gw}(f)\approx 10^{-15}$ across many orders of magnitude in frequency $f$~\cite{maggiore}.
Here, $d\rho_\text{gw}$ is the gravitational-wave energy density between $f$ and $f+df$ while $\rho_c$ is the critical energy density for a flat universe.

Such a low value of $\Omega_\text{gw}$ is unlikely to be directly detected by all but the most ambitious space-based gravitational-wave detectors~\cite{bbo,decigo}.
However, in models with either non-standard inflation or non-standard cosmology, it is possible to generate inflationary backgrounds accessible by current detectors~\cite{alterntive_cosmologies}.
Alternatively, it may be possible for the inflaton to decay non-perturbatively through parametric resonance.
This process, known as preheating, may produce a potentially detectable  gravitational-wave background through explosive particle production, peaking as high as $\Omega_\text{gw}\approx 10^{-11}$~\cite{preheating1, preheating2}.
In reality, the physics of inflation is highly uncertain.
Indirect detection, via the observation of $B$-modes in the cosmic microwave background, provides an alternative means of observing inflationary gravitational waves~\cite{cmb}.

Phase transitions in the early Universe may produce gravitational waves if they are strongly first-order~\cite{Kamionkowski:1993fg, Kahniashvili:2008pf, Caprini:2009fx, Caprini:2015zlo}.
The peak frequency of the gravitational-wave energy density spectrum $f_0$ is related to the energy scale of the transition $T_*$~\cite{maggiore,Kahniashvili:2008pf,Allen:1996}:
\begin{align}
f_0 \approx \unit[170]{Hz}\left(\frac{T_*}{\unit[10^9]{GeV}}\right) .
\end{align}
Thus, the detection of a primordial background from a phase transition by either an audio-band or millihertz gravitational-wave detector, such as LIGO~\cite{aLIGO} or LISA~\cite{lisa}, respectively, would probe physics at energy scales inaccessible by colliders, corresponding to a time when the Universe was only $\gtrsim\unit[10^{-14}]{s}$ old.
The energy density created from phase transitions depends on model-dependent details, but numerical simulations and scaling arguments suggest that $\Omega_\text{gw}(f_0)\approx 10^{-12\pm2}$ for a strongly first-order transition~\cite{thumb}.
This is just below the projected sensitivity of advanced detectors operating at design sensitivity~\cite{aLIGO, TheVirgo:2014hva}, but well within the range of space-based detectors and proposed third-generation terrestrial detectors~\cite{locus,Reitze:2019iox,ET}.

Astrophysical foregrounds are interesting in their own right since they contain valuable information about the population properties of compact binaries at high redshifts~\cite{Mandic2012, Vitale:2018yhm, Callister2020, Safarzadeh:2020qru}.
However, recent observations of merging compact binaries~\cite{Abbott:2016blz, TheLIGOScientific:2017qsa, LIGOScientific:2018mvr, Abbott:2020uma, LIGOScientific:2020stg, Abbott:2020khf} by the Advanced LIGO~\cite{aLIGO} and Virgo~\cite{TheVirgo:2014hva} detectors imply that primordial backgrounds are masked by much brighter astrophysical foregrounds~\cite{gw150914_stoch,gw170817_stoch}.
Binary black holes (BBHs) and binary neutron stars (BNSs) each produce astrophysical foregrounds of $\Omega_\text{gw}(f=\unit[25]{Hz})\approx 10^{-9}$ with $\alpha=2/3$~\cite{Zhu:2011bd, Wu:2011ac, Rosado_2011, Zhu:2012xw, gw170817_stoch}.
Some fraction of these astrophysical foregrounds is {\em resolvable} with current detectors, meaning that some of the events contributing to the background are unambiguously detectable.
The most ambitious proposed detectors will resolve essentially every compact binary in the visible Universe~\cite{bbo,decigo, DiggingDeeper}. 
These astrophysical foregrounds are non-Gaussian because the signals do not combine to create a random signal, characterized only by its statistical properties. 
Rather, BBHs merge every $\unit[2-10]{minutes}$ while BNSs merge every $\unit[4-62]{s}$~\cite{gw170817_stoch}.
While there are likely to be many BNS signals in the LIGO/Virgo band at any given time, they are nonetheless distinguishable based on their different coalescence times~\cite{DiggingDeeper,Vivanco2019}.

Previous proposals to disentangle the primordial background from the astrophysical foreground utilize the concept of subtraction.
The idea, pioneered in~\cite{cutler-harms}, is to measure the parameters of each resolved compact binary in order to subtract the gravitational waveform from the data.
Inevitably, this results in ``residuals,'' or systematic error from imperfect subtraction.
However, the residuals can be ``projected out'' using a Fisher matrix formalism, which reduces the level of contamination~\cite{Harms:2008xv}.
While~\cite{cutler-harms} considered the subtraction problem in the context of the ambitious Big Bang Observer~\cite{bbo}, more recent work has explored the possibility of carrying out subtraction using the third-generation detectors Einstein Telescope~\cite{Punturo:2010zz} and Cosmic Explorer~\cite{Reitze:2019iox} that are planned to come online in the next decade~\cite{DiggingDeeper, Sharma:2020btq}.

One limitation of the subtraction paradigm is that weak, unresolved signals are not subtracted and therefore contaminate the measurement of the primordial background, introducing a systematic error. While BBH mergers will be more easily resolvable, sub-threshold BNS mergers will impact the sensitivity of third-generation ground-based gravitational-wave detectors to the primordial background~\cite{Sachdev:2020bkk}.

Other analyses have proposed methods for the simultaneous measurement of stochastic gravitational-wave backgrounds with different spectral shapes~\cite{Ungarelli_2004, Parida_2016}. However, none of these methods account for the non-Gaussianity of the astrophysical foreground, resulting in a decrease in the sensitivity of the search. 

Here, we present a Bayesian formulation in which the primordial background and the astrophysical foreground are measured simultaneously.
Our method estimates the astrophysical foreground from both resolved and unresolved binaries, which ensures that our measurement of the primordial background is free from bias. 
The method can therefore also include the contributions from high signal-to-noise ratio compact binaries.
Because our likelihood models the non-Gaussianity of the astrophysical foreground as in \cite{tbs}, this method serves as the scaffolding for a unified, statistically optimal approach (yielding the minimum unbiased credible interval posterior) to the simultaneous detection of compact binaries and the primordial background.

{\em Formalism.}---We seek to measure a cosmological stochastic background described by two power-law parameters:
\begin{align}
\Omega_\text{gw}(f) = \Omega_\alpha \left(\frac{f}{\unit[25]{Hz}}\right)^{\alpha}.
\end{align}
Here, $\alpha$ is a power-law index while $\Omega_\alpha$ is the amplitude. The power-law model is chosen for consistency with cross-correlation searches for the stochastic background (e.g.~\cite{allen_romano, gw150914_stoch, gw170817_stoch, RomanoCornish}), but the subsequent formalism can be applied to any spectral shape.
The background is obscured by a foreground of merging compact binaries, each described by a vector of fifteen parameters $\theta$ including properties such as the component masses and the sky location.
We only consider BBH mergers for this analysis and assume that the population distribution for the binary parameters $\pi(\theta)$ is known to curtail the computational cost and additional complications for longer-duration BNS signals, but later discuss how the method can be generalized to relax these assumptions.
Since we want our formalism to include sub-threshold events, the number of compact binary signals in the data is, by assumption, unknown.

Following \cite{RomanoCornish} and \cite{tbs}, the likelihood of observing frequency-domain strain data, $s_{i,k}$, with a Gaussian stochastic background characterized by the parameters $(\Omega_{\alpha}, \alpha)$ and a compact binary coalescence with signal $h_{k}(\theta)$ is derived by marginalizing over the random Gaussian strain perturbation of the background:
\begin{widetext}
\begin{align}\label{eq:L}
&{\cal L}(s_{i,k}|\theta,\Omega_\alpha,\alpha) = \frac{1}{\det(\pi T \mathbf{C}_{k}(\Omega_\alpha,\alpha)/2)} \exp{\left(-\frac{2}{T}\left(s_{i,k}-h_{k}(\theta)\right)^{\dagger} \mathbf{C}^{-1}_{k}(\Omega_{\alpha},\alpha)\left(s_{i,k}-h_{k}(\theta)\right)\right)},
\end{align}
\end{widetext}
Here, we assume that the data is divided into segments of duration $T$ labeled with index $i$. The frequency dependence is denoted with the index $k$ such that $s_{i,k} = s_{i}(f_{k})$. The strain data in each segment, $s_{i,k}$, and the binary signal model, $h_{k}(\theta)$, are vectors with one entry for each detector in some network:
\begin{align}
s_{i,k} =  \left( \begin{array}{c} s^{(1)}_{i,k} \\ s^{(2)}_{i,k} \end{array} \right) \quad h_{k}(\theta) =  \left( \begin{array}{c} h^{(1)}_{k}(\theta) \\ h^{(2)}_{k}(\theta) \end{array} \right).
\end{align} 
At least two detectors are required to search for stochastic backgrounds modeled as excess cross-power, since the auto-power of one detector cannot distinguish between instrumental noise and signal, but the framework presented here can be extended to include multiple detector baselines.

The frequency-dependent covariance matrix, $\mathbf{C}_{k}$, includes contributions from both the detector noise power spectral density (PSD) $P_{I}(f_{k})$ and the primordial background energy density:
\begin{align}
\label{eq:C}
{\bf{C}}_{k} = & \left( \begin{array}{cc} P_{1}(f_{k}) + \kappa_{11}(f_{k})\Omega_\mathrm{gw} & \kappa_{12}(f_{k})\Omega_\mathrm{gw} \\ \kappa_{21}(f_{k})\Omega_\mathrm{gw} & P_{2}(f_{k}) + \kappa_{22}(f_{k})\Omega_\mathrm{gw} \end{array} \right),
\end{align}
where 
\begin{align}
\kappa_{IJ}(f_{k}) \equiv \gamma_{IJ}(f_{k}) \frac{3 H_0^2}{10\pi^2 f^3}
\end{align}
converts the primordial background energy density $\Omega_{\mathrm{gw}}$ into a (signal) strain power spectral density~\cite{allen_romano, Mingarelli:2019mvk}. The variable $\gamma_{IJ}(f_{k})$ is the overlap reduction function for detector pair $IJ$, encoding the geometry of the detector network~\cite{Christensen:1992wi, flanagan, RomanoCornish}.
It is normalized to $\gamma_{II}=1$ for coincident and coaligned detectors with perpendicular arms. 
Additionally, $H_0$ is the Hubble constant.
Combining data from many frequency bins, the likelihood is the product of the individual-frequency likelihoods: 
\begin{align}
    \mathcal{L}(s_{i}|\theta, \Omega_{\alpha}, \alpha) = \prod_{k}^{m}\mathcal{L}(s_{i,k}|\theta, \Omega_{\alpha}, \alpha).
\end{align}

For an astrophysical non-Gaussian foreground, we are interested in determining the fraction of segments containing a signal, $\xi$, (which we call the ``duty cycle'' following~\cite{tbs}) rather than the binary parameters, $\theta$, for a particular segment.
We say that a segment ``contains'' a binary signal if the time of coalescence falls inside the segment. 
In this case, the likelihood in Eq.~\ref{eq:L} can be marginalized over the binary parameters $\theta$ to obtain
\begin{align}\label{eq:xi}
{\cal L}\left(s_{i}|\Omega_\alpha, \alpha, \xi \right) = 
\xi{\cal L}_{S}(s_{i} | \Omega_\alpha,\alpha) + (1-\xi){\cal L}_{N}(s_{i} | \Omega_\alpha,\alpha),
\end{align}
where we have defined the marginalized signal and ``noise'' likelihoods as:
\begin{align}
\label{eq:ZS}
{\cal L}_S(s_{i}\vert \Omega_\alpha,\alpha)
= & \int d\theta \,
{\cal L}(s_i|\theta,\Omega_\alpha,\alpha) \, \pi(\theta) \\
\label{eq:ZN}
{\cal L}_N(s_{i}\vert\Omega_\alpha,\alpha) = 
& {\cal L}(s_i|\theta=0,\Omega_\alpha,\alpha).
\end{align}
The $\theta=0$ appearing in the expression for ${\cal L}_N$ indicates that the noise likelihood is functionally identical to the signal likelihood if we set the compact binary signal, $h_{k}(\theta)$, equal to zero. 
Readers should understand the phrase ``noise likelihood'' to refer to noise + a low-level Gaussian stochastic background but no binary signal.
We assume that the probability of observing one BBH merger event in a single segment is much less than one, so that the probability of observing two events is negligibly small, which is a reasonable assumption for BBH mergers~\cite{gw170817_stoch, tbs, DiggingDeeper}. 
We discuss how this assumption can be relaxed later. 

For an ensemble of $N$ data segments, $\{s\}$, the total likelihood is given by multiplying the likelihoods for individual segments:
\begin{align}\label{eq:tot}
{\cal L}\left(\{s\}|\Omega_\alpha, \alpha, \xi \right) = 
\prod_i^N \mathcal{L}(s_{i} | \Omega_{\alpha}, \alpha, \xi).
\end{align}
This joint likelihood function for $(\Omega_\alpha,\alpha,\xi)$ defined in Eq.~\ref{eq:tot} is the product of $N$ single-segment likelihood functions, each of which contains a signal sub-hypothesis (with probability $\xi$) and a noise sub-hypothesis (with probability $1-\xi)$.

To obtain joint posteriors on $(\Omega_\alpha, \alpha, \xi)$, we apply Bayes theorem:
\begin{align}
    p(\Omega_{\alpha}, \alpha, \xi| \{s\}) = \frac{\pi(\Omega_{\alpha}, \alpha, \xi)}{\mathcal{Z}}\mathcal{L}(\{s\} | \Omega_{\alpha}, \alpha, \xi),
\end{align}
where $\mathcal{Z}$ is the Bayesian evidence given by marginalizing the total likelihood over the stochastic parameters,
\begin{align}
\label{eq:Ztot}
    \mathcal{Z} = & \int d\Omega_{\alpha} \, d\alpha\, d\xi\,
{\cal L}(\{s\}|\Omega_\alpha,\alpha, \xi) \, \pi( \Omega_{\alpha}, \alpha, \xi),
\end{align}
and $\pi( \Omega_{\alpha}, \alpha, \xi)$ is the prior.

\emph{Demonstration.}---
We demonstrate this formalism with mock data.
Assuming a two-detector network of the LIGO Hanford and Livingston observatories operating at design sensitivity~\cite{aLIGO}, we simulate data for $101$ non-overlapping segments each with a duration of $\unit[4]{s}$.
Each segment contains uncorrelated Gaussian noise\footnote{While terrestrial detectors may suffer correlated noise from Schumann resonances~\cite{Thrane:2013npa}, studies suggest that this noise can be subtracted with Wiener filtering~\cite{Thrane:2014yza}.} colored by the noise PSD $P(f_{k})$ of the interferometers as well as correlated Gaussian noise colored by the signal power spectral density of the primordial background. 
The correlated noise is simulated such that the cross power spectral density is given by $\kappa_{IJ}(f_{k})\Omega_{\mathrm{gw}}(f_{k})$
for a cosmological background characterized by $(\log{\Omega_\alpha}=-6, \alpha=0)$, where we use $\log \equiv \log_{10}$ throughout. While this amplitude is several orders of magnitude higher than that expected for primordial backgrounds, we have chosen this value so that our simulated cosmological signal corresponds to an unambiguous primordial-background detection with signal-to-noise ratio (SNR) of $\sim 5.4$ for 404 seconds of data observed with advanced LIGO. The value of $\alpha\approx 0$ is expected for the background due to slow-roll inflation~\cite{Christensen:2018iqi}.

Next, we randomly assign BBH mergers to 11 of our simulated segments for a corresponding duty cycle of $\xi = 11/101=0.11$. This duty cycle is higher than would be expected based on the current estimates of the BBH merger rate~\cite{gw170817_stoch}, but is just chosen for the purposes of our demonstration. The chirp mass, 
\begin{align}
    \mathcal{M} = \frac{(m_{1}m_{2})^{3/5}}{(m_{1}+m_{2})^{1/5}},
\end{align}
is drawn from a uniform prior over the range $(13, 45)~M_{\odot}$. 
The prior for the symmetric mass ratio,
\begin{align}
    \eta = \frac{m_{1}m_{2}}{(m_{1}+m_{2})^{2}},
\end{align}
is uniform over $(0.09876, 0.25)$. The sky locations and component spin orientations are distributed isotropically, with spin magnitudes ranging uniformly from 0 to 0.8, and the luminosity distance prior is $\propto d_{L}^{2}$ between 500 and 5000 Mpc. This results in a range of network optimal SNRs between 2.06 and 12.17 with a median of 3.54. Only the signal with the highest SNR corresponds to a confident detection. The rest of the simulated events have network optimal SNR $< 7$ that would not be individually detected with high confidence.

If we were using real LIGO data instead of simulated Gaussian noise, the noise power spectral density in the covariance matrix in Eq.~\ref{eq:C} would have to be estimated from the data itself. Estimates of the PSD include both of the terms on the diagonal of the covariance matrix, $P(f_{k})+\kappa_{II}(f_{k})\Omega_{\mathrm{gw}}$, since auto-power due to detector noise cannot typically be distinguished from the persistent Gaussian background~\cite{RomanoCornish}.
This results in a decrease in the sensitivity of the search, which we mimic in our demonstration by fixing the diagonal terms to the sum of the known noise PSD and the signal power from the simulated cosmological background. Hence, the diagonal terms do not contribute to the estimation of the $(\Omega_{\alpha}, \alpha)$ parameters, although simultaneously fitting a parameterized PSD model as in \cite{Cornish:2014kda} could be a possible future extension.

Evaluating the likelihood in Eq.~\ref{eq:tot} poses a computational challenge due to the product over $N$ single-segment likelihoods.
To overcome this issue, we use likelihood reweighting~\cite{Payne:2019wmy} to evaluate the marginalized signal likelihood (Eq.~\ref{eq:ZS}) and the noise likelihood (Eq.~\ref{eq:ZN}) on a grid in $(\Omega_\alpha, \alpha)$.
For each segment we use the \texttt{cpnest}~\cite{cpnest} nested sampler as implemented in the \texttt{Bilby}~\cite{bilby, Romero-Shaw:2020owr} package to obtain posterior samples for the binary parameters using the likelihood in Eq.~\ref{eq:L} under the assumption that there is no Gaussian background present: $\Omega_{\alpha} = 0$. The priors for the binary parameters are the same as those used to generate the BBH injections previously described. We use the IMRPhenomPv2 waveform model~\cite{Hannam:2013oca, Husa:2015iqa, Khan:2015jqa} for the compact binary signal, $h_{k}(\theta)$, in both the simulation and recovery.

The marginalized signal likelihood for each segment at a particular value of $(\Omega_\alpha, \alpha)$ is calculated via a Monte Carlo integral over the $n$ posterior samples obtained in the original sampling step:
\begin{align}
\label{eq:weights}
    \mathcal{L}_{S}(s_{i} | \Omega_{\alpha}, \alpha) = \frac{\mathcal{Z}_{0,i}}{n}\sum_{j}^{n} \frac{\mathcal{L}(s_{i} | \theta_{j}, \Omega_{\alpha}, \alpha)}{\mathcal{L}(s_{i} | \theta_{j}, \Omega_{\alpha}=0)},
\end{align}
where $\mathcal{Z}_{0,i}$ is the evidence calculated by the sampler using the likelihood where $\Omega_{\alpha}=0$:
\begin{align}
    \mathcal{Z}_{0,i} = \int  d\theta\, \mathcal{L}(s_{i}|\theta, \Omega_{\alpha}=0)\pi(\theta).
\end{align}
The noise likelihood in Eq.~\ref{eq:ZN} can be directly evaluated on the same grid in $(\Omega_\alpha, \alpha)$ as the reweighted signal likelihood. 
We use a $50 \times 50$ grid ranging from $\log{\Omega_{\alpha}} \in [-8, -4]$ and $\alpha \in [0, 4]$. 

\begin{figure}
\centering
\centerline{\includegraphics[width=0.5\textwidth]{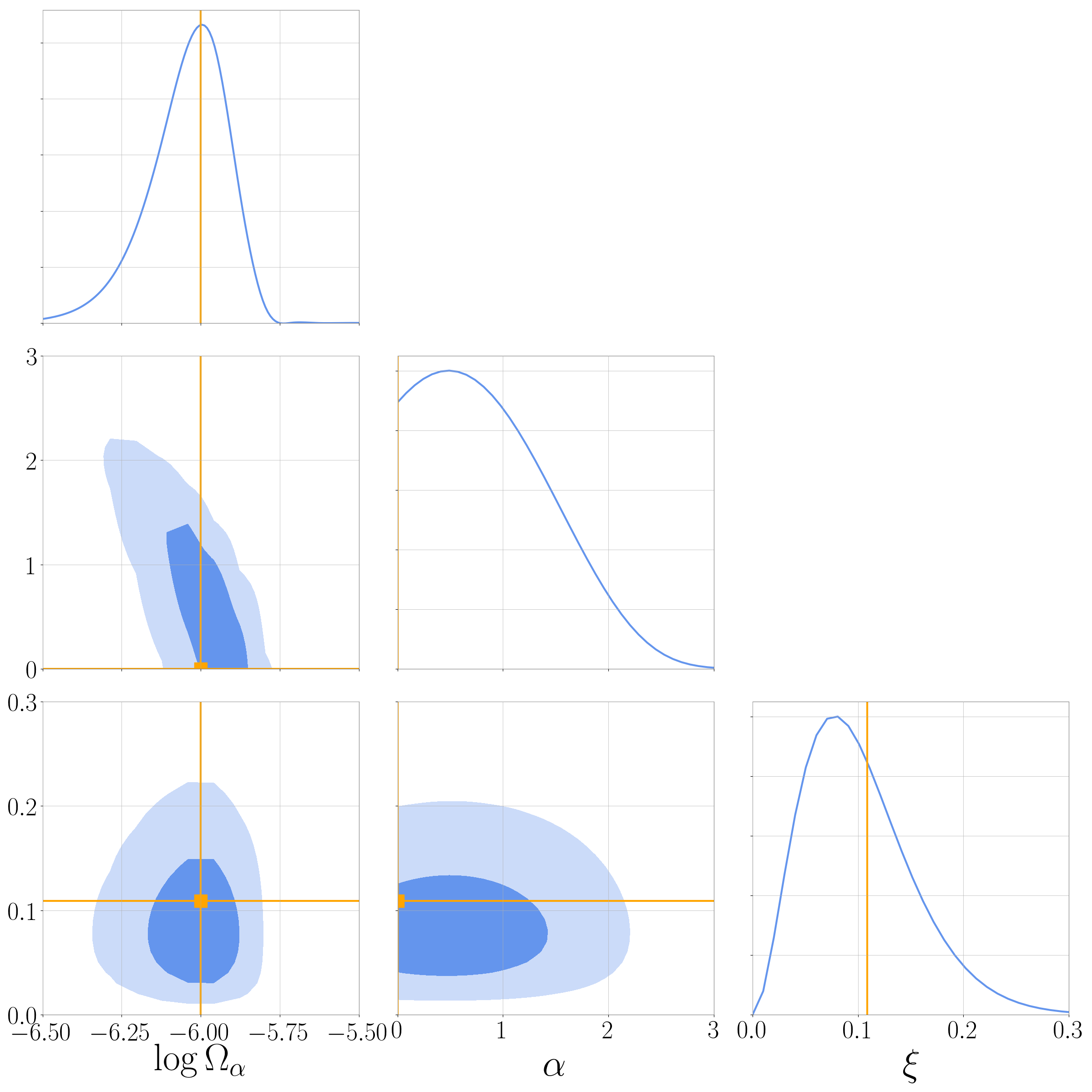}}
\caption{Corner plot for the combined posterior for $(\Omega_{\alpha}, \alpha, \xi)$, with the orange lines showing the true values used in the simulated data. The 90\% credible region is shown in light blue and the 50\% credible region in dark blue.}
\label{fig:corner_plot}
\end{figure}

Once we have obtained the marginalized signal and noise likelihoods for each segment using reweighting, we calculate the joint likelihood in Eq.~\ref{eq:xi} on a grid in $\xi$, with 100 values ranging from [0,1]. The full likelihood in Eq.~\ref{eq:tot} is then calculated by multiplying the individual three-dimensional grids from each data segment. Fig.~\ref{fig:corner_plot} shows the marginalized likelihoods for the cosmological background parameters $(\Omega_\alpha, \alpha)$ as well as $\xi$ obtained using all 101 simulated data segments. We recover values for all three parameters that are consistent with the true values used in the simulation: $\log{\Omega_{\alpha}} = -5.96_{-0.16}^{+0.08}$, $\alpha=0.49_{-0.49}^{+1.14}$, and $\xi=0.08_{-0.05}^{+0.09}$, where the uncertainty is the 90\% credible interval calculated using the highest probability density method. 

In addition to successfully measuring the parameters characterizing both the astrophysical foreground and the cosmological stochastic background simultaneously, we also calculate a Bayes factor comparing the cosmological signal hypothesis to the no-signal hypothesis. This quantifies to what extent the model where $\Omega_{\alpha}=0$ is statistically disfavored compared to the model where $(\Omega_{\alpha}, \alpha)$ can take on any of the values on our grid.
In the high-SNR limit, the natural log of the Bayes factor is proportional to the square of the SNR familiar from frequentist cross-correlation searches, $\ln{\mathrm{BF}^{S}_{N}} \sim \mathrm{SNR}^{2}/2$ (see, e.g.,~\cite{Thrane:2018qnx}). The ``signal'' evidence for a non-zero cosmological background is given by Eq.~\ref{eq:Ztot}. 
We set the priors on $\alpha$ and $\log{\Omega_{\alpha}}$ to be uniform across the ranges covered by the grid.
The ``noise'' evidence is evaluated by integrating Eq.~\ref{eq:tot} assuming that $\Omega_{\alpha}=0$:
\begin{align}
    \mathcal{Z}_{N} = \int d\xi \prod_{i}\xi \mathcal{Z}_{0,i}+(1-\xi)\mathcal{Z}_{N,i},
\end{align}
where $\mathcal{Z}_{N,i}$ is the likelihood in Eq.~\ref{eq:L} evaluated with {\em both} $\Omega_{\alpha}=0$ and $h_{k}(\theta)=0$. 
We obtain $\ln{\mathrm{BF}^{S}_{N}} = \ln{\mathcal{Z}_{S}} - \ln{\mathcal{Z}_{N}} = 11.16$, which is consistent with the naive scaling based on SNR for a signal with $\mathrm{SNR}=5.41$.

\emph{Discussion.}---In this paper we have demonstrated a new method for simultaneously detecting two distinct stochastic gravitational-wave backgrounds---a non-Gaussian astrophysical foreground from sub-threshold merging BBHs and a Gaussian cosmological background. Our method models both contributions simultaneously, so that subtraction of the foreground is not required. Additionally, this is the statistically optimal framework for detecting a stochastic background consisting of both a Gaussian and non-Gaussian component, resulting in significant improvements in the estimated time-to-detection of the astrophysical foreground compared to other methods for multi-component analyses, as described in \cite{tbs}. However, in the absence of a non-Gaussian foreground, we find that there is no statistical advantage to using the fully Bayesian method compared to the standard cross-correlation method. Based on the comparison of the signal-to-noise Bayes factor and SNR for the presence of the Gaussian background, the two methods yield a similar level of statistical confidence, to the extent that it is possible to compare frequentist and Bayesian detection statistics.

In our demonstration, we assume that the sampling priors chosen for the BBH parameters, $\pi(\theta)$, match the true population distribution. In order to avoid biases that would be introduced due to a mismatch between the population distribution and the sampling prior, our method could be amended to instead measure these population priors simultaneously with the cosmological background parameters and duty cycle, following the formalism described in \cite{tbs_populations}. This would amount to adding additional hyper-parameters to the marginalized signal likelihood in Eq.~\ref{eq:weights}:
\begin{align}
\label{eq:hyper}
    \mathcal{L}_{S}(s_{i} | \Lambda, \Omega_{\alpha}, \alpha) = \frac{\mathcal{Z}_{0}}{n}\sum_{j}^{n}\frac{\mathcal{L}(s_{i} | \theta_{j}, \Omega_{\alpha}, \alpha)\pi(\theta_{j} | \Lambda)}{\mathcal{L}(s_{i} | \theta_{j}, \Omega_{\alpha}=0, \alpha)\pi_{0}(\theta_{j})}.
\end{align}
The hyper-parameters $\Lambda$ describe the shape of the distribution $\pi(\theta | \Lambda)$, while the original prior used in the first step of sampling, $\pi_{0}(\theta)$, must also be divided out. The hyper-parameters do not enter the noise likelihood in Eq.~\ref{eq:ZN} because the noise model assumes that each segment contains only the cosmological background with no binary signal.

Evaluating the marginalized signal likelihood in Eq.~\ref{eq:hyper} using the same grid-based reweighting technique becomes computationally prohibitive, since the hyper-parameters $\Lambda$ drastically increase the dimensionality of the grid. 
One possible solution that has been applied to similar problems in gravitational-wave astronomy could be to build a high-dimensional interpolant~\cite{Vivanco:2019qnt, Wysocki:2020myz}.
Another promising approach could be to factorize the problem into two separate calculations, first carrying out population studies ignoring the stochastic background then using the inferred posterior predictive distributions for $\pi(\theta | \Lambda)$ as priors for the $\Omega>0$ run.
We leave exploration of these approaches to future work.

Another simplifying assumption we make in our demonstration is that only merging BBHs contribute to the astrophysical foreground, while in reality there will also be a foreground from binary neutron star and neutron star-black hole mergers. While our assumption that there is only one binary signal
in a $\unit[4]{s}$ analysis segment is valid for BBHs, the rate of BNS mergers is higher, meaning that there are typically $\sim15$ unresolved BNS signals in the LIGO band at any given time~\cite{gw170817_stoch}. 


Because we need to model multiple populations of merging binaries simultaneously to avoid contamination from residual power, 
one possible solution would be to treat the number of binary mergers in a given segment as a free parameter using a trans-dimensional Markov Chain Monte Carlo algorithm, fitting the binary parameters for multiple mergers along with the cosmological background parameters all at once~\cite{tdmcmc}.
Another possible method is to analyze overlapping stretches of data that are offset by a shorter $\unit[0.2]{s}$ window, constraining the coalescence time prior to this window so that at most one binary system merges during this short ``segment'', allowing us to keep the same definition of $\xi$ presented above.
Preliminary tests suggests that the presence of other binary signals, merging at times outside of the segment, have a negligible effect on inferences about the binary merging during the segment.
By marginalizing over the BNS parameters in many short segments, it should be possible to calculate the likelihood of a much longer span of data given the stochastic parameters. We estimate that it would take about $\sim 10^{5}$ CPUs to perform the BNS analysis in real time~\cite{Finstad:2020sok}, followed by $\sim 10$ GPUs to perform the hierarchical inference including the uncertainty in the population distribution using the likelihood interpolation method for each individual segment~\cite{Talbot:2019okv}.
We leave investigation of these methods to future work.

The formalism we describe and demonstrate assumes that the uncorrelated detector noise is Gaussian, while it is known that interferometric gravitational-wave data suffers from non-Gaussian noise transients called glitches~\cite{TheLIGOScientific:2016zmo}. This assumption can be relaxed via the introduction of additional duty cycle parameters to the likelihood in Eq.~\ref{eq:L}, characterizing the fraction of segments that contain a glitch in each detector, as described in \cite{tbs}. This would increase the computational cost for each data segment analyzed, but the method is embarrassingly parallelizable, so the overall wall time for running the analysis does not increase significantly.

We also note that limitations in the accuracy of the waveform model describing the compact binary signal can leave behind coherent residual power that could bias the inference of the Gaussian background parameters. 
Based on current estimates of the uncertainty in numerical relativity waveforms~\cite{Purrer:2019jcp}, this level of contamination would likely not affect cosmological backgrounds probeable with proposed third-generation detectors, but improvements to waveform modeling would be necessary to recover unbiased parameter estimates for the weakest background models. The subtraction-projection methods for background detection would also be affected by waveform systematics, but our method could be modified to account for marginalizing over different waveform models~\cite{Ashton:2019leq} or parameterizing the waveform uncertainty~\cite{Edelman:2020aqj}.

While we demonstrate our method for the simultaneous detection of a stochastic background with both Gaussian and non-Gaussian components in the context of a cosmological background and an astrophysical foreground of BBH mergers, this formalism can be applied to any analogous problem. 
For example, this method could be applied to simultaneously measure both individual compact binary mergers or a foreground of these sources in the frequency band of the space-based LISA detector~\cite{lisa} on top of the white dwarf confusion noise background~\cite{Bender_1997, Adams_2014}. Our model can also be extended to include multiple Gaussian backgrounds with different spectral shapes through the addition of extra terms in the covariance matrix defined in Eq.~\ref{eq:C}. One such example is the contamination from correlated magnetic noise in a ground-based detector network~\cite{Christensen:1992wi, Thrane:2013npa, Thrane:2014yza}, which has a unique overlap reduction function~\cite{Himemoto:2017gnw}.  

\begin{acknowledgments}
\emph{Acknowledgements.}---The authors thank Jan Harms and Salvatore Vitale for thoughtful comments on the manuscript. SB and CT acknowledge support of the National Science Foundation and the LIGO Laboratory. 
LIGO was constructed by the California Institute of Technology and Massachusetts Institute of Technology with funding from the National Science Foundation and operates under cooperative agreement PHY-1764464.
SB is also supported by the Paul and Daisy Soros Fellowship for New Americans, the Australian-American Fulbright Commission, and the NSF Graduate Research Fellowship under Grant No. DGE-1122374.
ET and RS are supported by the Australian Research Council (ARC) CE170100004. ET is supported by ARC FT150100281.
The authors are grateful for computation resources provided by the OzStar cluster.
This article carries LIGO Document Number LIGO-P2000297.
\end{acknowledgments}

\bibliography{cosmo}

\end{document}